\documentclass[prl,aps,twocolumn,showpacs]{revtex4}
\usepackage{epsfig}
\def\Xint#1{\mathchoice
   {\XXint\displaystyle\textstyle{#1}}%
   {\XXint\textstyle\scriptstyle{#1}}%
   {\XXint\scriptstyle\scriptscriptstyle{#1}}%
   {\XXint\scriptscriptstyle\scriptscriptstyle{#1}}%
   \!\int}
\def\XXint#1#2#3{{\setbox0=\hbox{$#1{#2#3}{\int}$}
     \vcenter{\hbox{$#2#3$}}\kern-.5\wd0}}

\def\dashint{\Xint-}
\include{graphics}
\begin{document}
\title{Self-similar relaxation dynamics of a fluid wedge in a Hele-Shaw cell}
\author{Omri Gat, Baruch Meerson, and Arkady Vilenkin}
\affiliation{$^{1}$Racah Institute  of  Physics, Hebrew University
of Jerusalem, Jerusalem 91904, Israel}
\begin{abstract}
Let the interface between two immiscible fluids in a Hele-Shaw cell
have, at $t=0$, a wedge shape. As a wedge is scale-free, the fluid
relaxation dynamics are self-similar. We find the dynamic exponent
of this self-similar flow and show that the interface shape is given
by the solution of an unusual inverse problem of potential theory.
We solve this problem analytically for an almost flat wedge, and
numerically otherwise. The wedge solution is useful for analysis of
pinch-off singularities.
\end{abstract}
 \maketitle

\textit{Introduction.} Interface dynamics between two immiscible fluids in a
Hele-Shaw cell have attracted a great interest in the last two decades. Most of
the efforts have dealt with forced flows, when a more viscous fluid is displaced
by a less viscous fluid. In the forced case the viscous fingering instability
\cite{ST,Paterson} develops and brings about intricate issues of pattern
selection in a channel geometry
\cite{
Kadanoff,Kessler,Casademunt}, development of fractal structure in a radial
geometry \cite{Paterson2}, \textit{etc}. The role of small surface tension in
the theory of a forced Hele-Shaw flow is to introduce regularization on small
scales. This Letter deals with an \textit{unforced} Hele-Shaw (UHS) flow
\cite{Constantin1,Almgren,Sharon,CLM,VMS}, where surface tension at the fluid
interface is the \textit{only} driving factor. The pertinent free boundary
problem here is non-integrable and, because of its non-locality, hard for
analysis. To our knowledge, the only known analytical solutions to this class of
problems are (i) a linear analysis of the dynamics of a slightly deformed flat
or circular interface \cite{ST,Paterson} and (ii) a recent asymptotic scaling
analysis of the dynamics of a long stripe of an inviscid fluid trapped in a
viscous fluid \cite{VMS}. To get more insight into the physics of UHS flows, we
address here the case when one of the fluids at $t=0$ has the form of a wedge.
In this case the flow is self-similar. Building on this simplification, we
recast the problem into an unusual inverse problem of potential theory. We solve
this problem analytically for an almost flat wedge and numerically for several
other wedge angles. Finally, we use a wedge solution for analysis of pinch-off
events of the UHS flow, which has attracted much interest in theory and
experiment \cite{Almgren,Sharon}.

\textit{Governing equations and self-similarity.} Let one of the fluids have a
negligible viscosity, so that the pressure inside this fluid is constant and can
be taken zero. The velocity of the viscous fluid is
$\mathbf{v}\,(\mathbf{r},t)=-(b^2/12\mu) \,\nabla p\,(\mathbf{r},t)$, where $p$
is the pressure, $\mu$ is the dynamic viscosity, and $b$ is the plate spacing
\cite{ST,Paterson,Kadanoff}. Therefore, the interface speed is
\begin{equation}\label{speed}
  v_n = - (b^2/12 \mu) \partial_n p\,,
\end{equation}
where index $n$ denotes the components of the vectors normal to the interface
outwards, and $\partial_n p$ is evaluated at the respective points of the
interface $\gamma$. As $\mathbf{\nabla} \cdot \mathbf{v}=0$ in the
(incompressible) viscous fluid, the pressure there is a harmonic function:
\begin{equation}\label{Laplace}
  \nabla^2 p =0\,.
\end{equation}
The Gibbs-Thomson relation at the interface yields
\begin{equation}\label{jump2}
  p\,|_{\gamma} = (\pi/4)\,\sigma {\cal K}\,,
\end{equation}
where $\sigma$ is surface tension, and ${\cal K}$ is the local
curvature of the interface, positive when the inviscid region is
convex outwards. As the flow is undriven we demand
\begin{equation}\label{external}
\partial_n p\, = 0 \;\;\;\mbox{at}\;\;\;\mathbf{r}\to \infty\,.
\end{equation}
We assume that the interface has the form of a graph $y=y(x,t)$ and
rewrite Eq. (\ref{speed}) as an evolution equation:
\begin{eqnarray}
 \nonumber  \partial_t y(x,t) &=& - (b^2/12 \mu)\,\partial_n p\, \sqrt{1+(\partial_x
    y)^2} \\
  &=& (b^2/12 \mu) \left[\partial_x y(x,t)\,\partial_x p-\partial_y
  p\right]\,,
  \label{interface}
\end{eqnarray}
where the derivatives of $p$ are evaluated at the interface.
\begin{figure}
\includegraphics[width=5.0 cm,clip=]{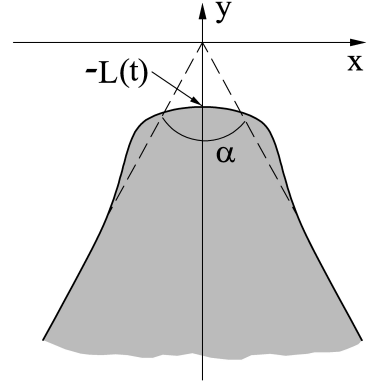}
\caption{The setting for fluid wedge relaxation.} \label{figwedge}
\end{figure}
At $t=0$ the inviscid fluid has the form of a wedge of angle
$\alpha$, so that $y=-|x| \cot\, (\alpha/2)$, see
Fig.~\ref{figwedge}. As this initial condition and Eqs.
(\ref{Laplace})-(\ref{interface}) do not introduce any length scale,
the solution must be self-similar \cite{B}. Let $L(t)$ be the
retreat distance of the wedge tip. Then the interface position and
the pressure in the viscous fluid can be written as
\begin{eqnarray}
\label{ss1} \label{interfaceSS}
  y(x,t) &=& L(t)\, \Phi \left[\frac{x}{L(t)}\right]\,, \\
  \label{pressureSS}
  p(x,y,t) &=& \frac{\pi \sigma}{4L(t)}\,P\left[\frac{x}{L(t)},\frac{y}{L(t)}\right]\,,
\end{eqnarray}
respectively. We fix the coordinates by choosing $y(x=0,t)=-L(t)$, that is
$\Phi(0)=-1$. In the rescaled coordinates $X=x/L(t)$ and $Y=y/L(t)$ the
Laplace's equation (\ref{Laplace}) keeps its form, while Eq.~(\ref{jump2})
becomes
\begin{equation}\label{dirichlet}
    P\left[ X,Y=\Phi(X)\right] = \frac{\Phi^{\prime \prime}(X)}{\left[
    1+\left[
    \Phi^{\prime}(X)\right]^2\right]^{3/2}}\,,
\end{equation}
where primes stand for $X$-derivatives. Now, using Eqs.
(\ref{interfaceSS}) and (\ref{pressureSS}) in Eq. (\ref{interface}),
we arrive at the following equation:
\begin{equation}\label{neumann}
    \Phi^{\prime}(X)\, \partial_XP-\partial_YP=\lambda \left[\Phi(X)-X\Phi^{\prime}(X)\right]\,,
\end{equation}
where the derivatives of $P$ are evaluated at the rescaled interface $\Phi(X)$,
$L(t)=\left[(\pi \lambda \sigma b^2 t)/(16 \mu)\right]^{1/3}$, and $\lambda$ is
an unknown dimensionless parameter. The boundary conditions are $\Phi(0)=-1,\,
\Phi^{\prime}(0)=0$ and $\Phi(X\to \pm\infty) = -|X| \cot\,(\alpha/2)$. Note
that we have already found the dynamic scaling exponent $1/3$: the same as
observed in the relaxation of fractal viscous fingering patterns
\cite{Sharon,CLM}. The shape function $\Phi (X)$ (and the parameter $\lambda$)
for a given wedge angle $\alpha$ is determined by the solution of the following
(quite unusual) inverse problem of potential theory. A harmonic function
$P(X,Y)$ must obey \textit{both} a Dirichlet boundary condition [Eq.
(\ref{dirichlet})], \textit{and} a Neumann boundary condition [Eq.
(\ref{neumann})], while the function $\Phi(X)$ must be determined from the
demand that these two conditions be consistent. We solved this problem
analytically for an almost flat wedge, and numerically otherwise. Before
reporting the analytic solution, we present a large-$X$ asymptote of $\Phi(X)$,
valid for \textit{any} wedge angle. It corresponds to the leading term of the
multipole expansion of $P(X,Y)$ at large distances. Introduce, for a moment,
polar coordinates $r, \phi$ with the origin at the point $X=Y=0$ and measure the
polar angle $\phi$ from the ray $Y=-X \cot\,(\alpha/2)$ counterclockwise. At
large $|X|$ the curve $Y=\Phi(X)$ is almost flat, so $P[X,\Phi(X)] \to 0$ there
by virtue of Eq. (\ref{dirichlet}). Therefore, the leading term of the multipole
expansion is $P(r\gg 1,\phi)=\mbox{const}\,r^{-\nu} \sin(\nu \phi)$, where
$\nu=(2-\alpha/\pi)^{-1}$ \cite{Jackson}. Now we employ Eq. (\ref{neumann}) and
obtain, at $|X|\gg 1$,
\begin{equation}\label{tail1}
    \Phi (X) =-|X| \cot\,(\alpha/2) 
    + C\, |X|^{-\frac{3 \pi-\alpha}{2\pi-\alpha}}+\dots\,,
\end{equation}
with an unknown constant $C$ that depends only on $\alpha$.

\textit{Almost flat wedge.} Let us assume that $\pi-\alpha \ll \pi$,
and introduce the small parameter $\varepsilon \equiv \cot
(\alpha/2)\ll 1$. We rescale the variables: $X=\xi/\varepsilon$,
$Y=\eta/\varepsilon$, $P(X,Y)=\varepsilon^2\,U(\xi,\eta)$ and
$\lambda=\Lambda \varepsilon^3$. In the rescaled varaibles the
interface equation is $\eta=\varepsilon \psi(\xi)$, where
$\psi(\xi)\equiv\Phi(\xi/\varepsilon)$. Keeping only leading terms,
we can rewrite the boundary conditions (\ref{dirichlet}) and
(\ref{neumann}) for the harmonic function $U(\xi,\eta)$ in the
following form:
\begin{eqnarray}
\label{first}
  U\left[ \xi,\varepsilon \psi(\xi)\right] &=& \psi^{\prime\prime}(\xi)\,, \\
  \label{second}
  \partial_{\eta}U \left[ \xi,\varepsilon \psi(\xi)\right]&=& \Lambda
  \left[\xi \psi^{\prime}(\xi)-\psi(\xi)\right]\,,
\end{eqnarray}
where
\begin{equation}\label{BCs}
\psi(0)=-1, \,\psi^{\prime}(0) = 0,\,\psi(\xi\to \pm\infty) = -|\xi|+ o(1)\,.
\end{equation}
The rescaled problem does not include $\varepsilon$, except in the second
argument of the functions on the left hand side of Eqs. (\ref{first}) and
(\ref{second}). In view of the condition $\psi(\xi\to \pm\infty) = -|\xi|$ one
cannot put the second argument to zero at sufficiently large $|\xi|$. As will be
shown below, these values of $\xi$ are \textit{exponentially} large in
$\varepsilon^{-1}$, while at shorter distances one can safely put the second
argument to zero.

The problem obtained in this way is soluble exactly. Assume
$\psi(\xi)$ is known. Then one can easily find the harmonic function
in the upper half-plane $\eta>0$, that satisfies the Dirichlet
condition $u (\xi,0) = \psi^{\prime\prime}(\xi)$ on the $\xi$-axis:
\begin{equation}\label{dirsol}
    U(\xi,\eta)=\frac{1}{\pi}\int_{-\infty}^{\infty}\frac{\eta \,\psi^{\prime\prime}
    (s)\,ds}{(\xi-s)^2+\eta^2}\,.
\end{equation}
Now we should impose the Neumann condition (\ref{second}) (where we
put $\varepsilon=0$). To avoid calculation of hyper-singular
integrals, we find the harmonic conjugate
\begin{equation}\label{conj}
    V(\xi,\eta)=\frac{1}{\pi}\int_{-\infty}^{\infty}\frac{(\xi-s) \psi^{\prime\prime}
    (s)\,ds}{(\xi-s)^2+\eta^2}
\end{equation}
and, by virtue of the Cauchy-Riemann conditions, replace
$\partial_{\eta}U(\xi,0)$ by $-\partial_{\xi}V(\xi,0)$. This yields
a non-standard integro-differential equation
\begin{equation}\label{intdiff}
    \Lambda
  \left[\xi \psi^{\prime}(\xi)-\psi(\xi)\right]= - \frac{1}{\pi}
  \frac{d}{d\xi} \,\dashint_{-\infty}^{\infty}\frac{\psi^{\prime\prime}
    (s)\,ds}{\xi-s}\,,
\end{equation}
where $\dashint$ denotes the principal value of the integral. Fortunately, upon
differentiation with respect to $\xi$ Eq. (\ref{intdiff}) becomes an equation
for $\psi^{\prime\prime}(\xi)$ which is soluble by Fourier transform. The result
is
\begin{equation}\label{secondder}
    \psi^{\prime\prime}(\xi)=-\frac{1}{\pi}\int_{-\infty}^{\infty}
    e^{-\frac{|k|^3}{3 \Lambda}}\,\cos k\xi\,dk
\end{equation}
(the constant of integration is determined from the condition
$\int^{\infty}_{-\infty}\psi^{\prime\prime}(\xi) d\xi=-2$).
Integrating twice in $\xi$ and using the first two conditions in Eq.
(\ref{BCs}) yields
\begin{equation}\label{phi1}
    \psi(\xi) = -1 -\frac{2}{\pi}\int_{-\infty}^{\infty}
    e^{-\frac{|k|^3}{3 \Lambda}}\,\,\frac{\sin^2\frac{k
    \xi}{2}}{k^2}\,dk\,.
\end{equation}
To determine $\Lambda$, we expand this expression at $|\xi| \gg 1$:
\begin{equation}\label{asymp1}
    \psi(\xi) = -|\xi|+\frac{2\,\Gamma(2/3)}{\pi
(3\Lambda)^{1/3}}-1+ \frac{2}{3 \pi \Lambda}\,\xi^{-2} + \dots\,,
\end{equation}
where $\Gamma(w)$ is the gamma-function \cite{singularity}. To
eliminate the offset ${\cal O}(1)$ we put $\Lambda=(8/3)\,
\pi^{-3}[\Gamma(2/3)]^3 =0.213545\dots$. Though the integrals in
Eqs. (\ref{secondder}) and (\ref{phi1}) can be expressed via the
generalized hypergeometric function $_pF_q(a;b;z)$, it is more
convenient to keep the integral form \cite{small}. To complete the
solution, we find the rescaled pressure:
\begin{equation}\label{pressure}
 U(\xi,\eta)=
 -\frac{1}{\pi}\int_{-\infty}^{\infty} e^{-\frac{|k|^3}{3 \Lambda}-|k|\eta}\cos k\xi\,\,dk\,.
\end{equation}

Now we find the distance $|\xi|=l(\varepsilon)\gg 1$ at which the
solution (\ref{phi1}) becomes inaccurate, and improve the
large-$|\xi|$ asymptote. Let us compare Eq. (\ref{asymp1}), which
becomes
\begin{equation}\label{tail2}
    \psi(\xi) = -|\xi|+ \frac{\pi^2}{4
[\Gamma(2/3)]^3}\,\xi^{-2}\,,\;\;1\ll |\xi| \ll l(\varepsilon)\,,
\end{equation}
with the large-$|\xi|$ multipole asymptote (\ref{tail1}):
\begin{equation}\label{tail3}
\psi (\xi) =-|\xi| +
C(\varepsilon)\,|\xi/\varepsilon|^{-\frac{3 \pi-\alpha}{2\pi-\alpha}}\,, \,\;\;|\xi|\gg 1\,.
\end{equation}
where, for small $\varepsilon$, $-(3\pi-\alpha)/(2\pi-\alpha)\simeq
-2+2\varepsilon/\pi\,.$ We see that the last term in Eq.
(\ref{tail2}) lacks the small correction $2\varepsilon/\pi$ in the
exponent of $\xi$.  We can match the two asymptotes (\ref{tail2})
and (\ref{tail3}) in their common region of validity $1\ll |\xi| \ll
l(\varepsilon)$. We define $l(\varepsilon)$ as the value of $|\xi|$
for which the correction to the exponent yields a factor $e$:
$l(\varepsilon)= e^{\pi/(2 \varepsilon)}$ [notice that, at
$|\xi|\sim l(\varepsilon)$, the deviation of $\psi(\xi)$ from its
flat asymptote $-|\xi|$ is already exponentially small: $\sim
e^{-\pi/\varepsilon}$]. The matching yields $C(\varepsilon)$, and we
arrive at the improved small-$\varepsilon$ large-$|\xi|$ asymptote:
\begin{equation}\label{tail4}
    \psi(\xi) = -|\xi|+ \frac{\pi^2}{4
[\Gamma(2/3)]^3}\,e^{-\frac{2 \varepsilon}{\pi}
\ln\varepsilon}\,\xi^{-2+\frac{2\varepsilon}{\pi}}\,.
\end{equation}

So far we have dealt with inviscid fluid wedges: $\alpha<180^\circ$.
Our results, however, can be immediately extended to viscous fluid
wedges: $\alpha>180^\circ$.

\textit{Numerical algorithm and parameters.}  For a general wedge angle the
shape function of the self-similar interface can be found numerically. Instead
of dealing with the similarity formulation of the problem
(\ref{dirichlet})-(\ref{neumann}), we computed the time-dependent relaxation of
wedges of different angles, as described by (rescaled) Eqs.
(\ref{speed})-(\ref{external}) \cite{units}. Our numerical algorithm \cite{VM}
employs a variant of the boundary integral method for an exterior Dirichlet
problem for a singly connected domain, and explicit tracking of the contour
nodes. The harmonic potential is represented as a superposition of potentials
created by a dipole distribution  with an unknown density $\mathbf{D}$ on the
contour. $\mathbf{D}$ is computed from a linear integral equation
\cite{Tikhonov}. Computing another integral of this dipole density yields the
harmonic conjugate, whose derivative along the contour is equal to the normal
velocity of the interface.

We chose the singly connected domain to be (i) a rhombus with angles $120^\circ$
and $60^\circ$, (ii) a square, and (iii) a straight cross with aspect ratio
$10^3$ \cite{units}. In this manner we could exploit the 4-fold symmetry of the
domains and measure the retreat distance of the respective vertexes, $L(t)$, and
the rescaled interface shapes $\Phi(X)$ for four wedge angles: $120^\circ$,
$90^\circ$, $60^\circ$ and $270^\circ$, the latter one corresponding to a
$90^\circ$ wedge of the \textit{viscous} fluid. The ultimate shapes of the
rhombus- and square-shaped domains are perfect circles. Therefore, to observe
the self-similar asymptotics we did the measurements at times much shorter than
the characteristic time of relaxation toward a circle, and at distances much
smaller than the domain size (so that the effect of the other vertexes could be
neglected). For the rhombus and square an equidistant grid with 901 nodes per
side was employed. For the quarter of the cross we used 2801 nodes. The time
step was taken to be $10^{-3}$ times the maximum of the ratio of the interface
curvature radius and the interface speed at the same node. The domain area
conservation was used for accuracy control. For the measurements reported here
the area was conserved with an accuracy better than $10^{-3}\%$.

\begin{figure}
\includegraphics[width=5.5 cm,clip=]{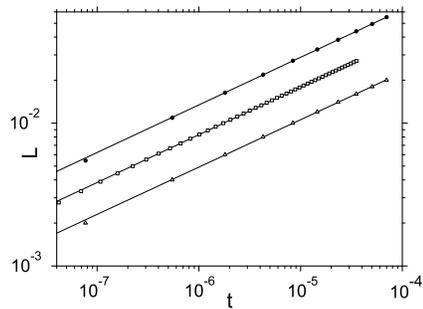}
\caption{The retreat distance $L(t)$ and its power-law fit for
$\alpha=120^\circ$ (triangles), $90^\circ$ (squares) and $60^\circ$
(circles).} \label{figretreat}
\end{figure}

\textit{Numerical results.} We first report the results for the three viscous
fluid wedges. Figure \ref{figretreat} shows the retreat distance $L(t)$ for the
angles $120^\circ$, $90^\circ$, and $60^\circ$. Power law fits yield $L(t)= 0.48
\, t^{0.33}$, $0.84 \, t^{0.33}$, and $1.33 \, t^{0.33}$, respectively, so the
dynamic exponent $1/3$ is clearly observed.  In the rescaled units, used in the
simulations \cite{units}, the analytical prediction for an almost flat wedge is
$L(t)= a t^{1/3}$, where $a=(3 \Lambda)^{1/3} \varepsilon \simeq 0.862 \,
\varepsilon$. For $\alpha=120^\circ$ and $90^\circ$ this yields $a \simeq 0.498$
and $a=0.862$, respectively, in very good agreement with the measured values
$0.48$ and $0.84$. Even for $\alpha=60^\circ$ the analytical prediction,
$a=1.493$, is only $12\%$ higher than the measured value $1.33$.

\begin{figure}
\includegraphics[width=7 cm,clip=]{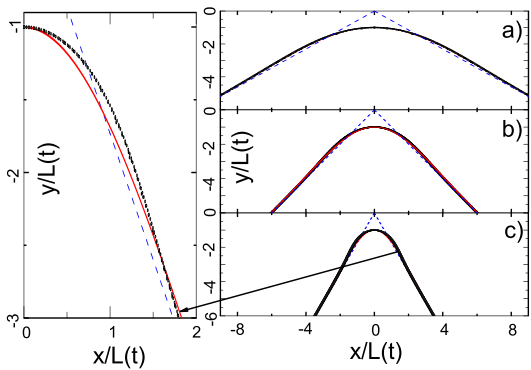}
\caption{(color online). Right panel: the shape function $\Phi$ 
for wedges of inviscid fluid: $\alpha=120^\circ$ (a), $90^\circ$ (b) and
$60^\circ$ (c). Data for three different times ($2.3 \times 10^{-7}$, $1.3
\times 10^{-5}$ and $1.1 \times 10^{-4}$ for a and c, and $8.2 \times 10^{-7}$,
$8.9 \times 10^{-6}$ and $3.5 \times 10^{-5}$ for b) collapse into a single
curve. The red solid line is the prediction of the almost-flat-wedge theory, the
blue dashed line is the asymptote $Y=-|X|\,\cot{\alpha/2}$. Left panel: a blowup
of a part of figure c.} \label{figshapes}
\end{figure}

The rescaled  shapes of the three evolving wedges are depicted in
Fig.~\ref{figshapes}. That the curves, measured at three different times,
collapse into a single curve proves self-similarity. The prediction of our
almost-flat-wedge theory, shown on the same three graphs, works very well for
$\alpha=120^\circ$ and $90^\circ$, and fairly well even for $60^\circ$.

We also measured, for each of the three values of the wedge angle, the tail of
the shape function (the difference between $\Phi(X)$ and
$Y=-|X|\,\cot{\alpha/2}$). The results are in excellent agreement with the
theoretical prediction, given by the last term in Eq.~(\ref{tail1}).

\begin{figure}
\includegraphics[width=6.0 cm,clip=]{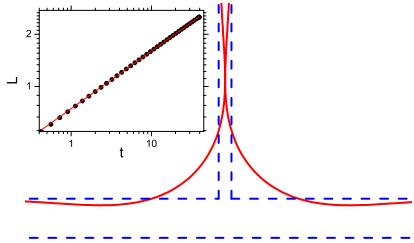}
\caption{(color online). Pinch-off of a straight branch of thickness $\Delta$ of
inviscid fluid. Each of the two viscous fluid wedges corresponds to
$\beta=90^{\circ}$ (\textit{i.e.} $\alpha=270^{\circ}$). The blue dashed lines:
the interface shape at $t=0$; the red solid lines: the interface shape at the
pinch-off time $t=50.65$ (in the units of \cite{units}). The inset: the measured
viscous fluid retreat distance versus time and its power-law fit $0.74 \times
t^{0.33}$.} \label{figbranch}
\end{figure}

\textit{Pinch-offs.} The self-similar wedge solutions are useful for analysis of
pinch-offs in UHS flows \cite{Almgren,Sharon}. Let the inviscid fluid domain
represent, at $t=0$,  an infinitely long straight branch, coming at an angle
$\beta$ from an infinitely long straight ``trunk". The simple physics in the
inviscid fluid branch precludes interaction between the two viscous fluid wedges
of angles $\beta$ and $\pi-\beta$, which evolve in a self-similar manner,
causing the inviscid branch to thin, and ultimately to pinch-off. The $t^{1/3}$
law intrinsic in the self-similar solution implies that the pinch-off time is
proportional to the branch thickness cubed. The interface shape at all times
prior to the pinch-off can be obtained, with a proper rescaling, from the
respective self-similar shape functions of the two viscous fluid wedges. The
case of $\beta=90^{\circ}$ is shown in Fig.~\ref{figbranch}, where the retreat
distance, the shape function and the pinch-off time are taken from the
previously described simulation of the cross-shaped domain with aspect ratio
$10^3$.

\textit{Summary.} We have studied analytically and numerically the surface
tension driven flow of a fluid wedge in a Hele-Shaw cell. We have shown that the
fluid interface evolves self-similarly, found the asymptotic interface shape at
large distances, and recast the problem into an unusual inverse problem of
potential theory. We solved this inverse problem analytically in the limit of
nearly flat wedge, and performed numerical simulation which support and extend
the analytic calculations. Like in the case of self-similar solutions, obtained
for wedge-like initial conditions in other surface tension driven flows
\cite{others}, this solution provides a sharp characterization of the UHS flow.
It also sheds a new light on the pinch-off singularities of this flow.


\begin{thebibliography}{111}
\bibitem{units} In the numerical simulations we measured the
distances in units of $\Delta$ (the side of the rhombus or square, the thickness
of the cross arm), the time in units of $48 \mu \Delta^3/(\pi \sigma b^2)$, and
the viscous fluid pressure in units of $\pi \sigma/(4\Delta)$. Then
Eqs.~(\ref{speed}) and (\ref{jump2}) become $v_n = - \partial_n p$ and
$p\,|_{\gamma} = {\cal K}$, so the rescaled problem is parameter-free.
\end{thebibliography}
\end{document}